\documentclass[conference]{IEEEtran}
\IEEEoverridecommandlockouts
\usepackage{cite}
\usepackage{amsmath,amssymb,amsfonts}
\usepackage{algorithmic}
\usepackage{graphicx}
\usepackage{textcomp}
\usepackage{xcolor}
\def\BibTeX{{\rm B\kern-.05em{\sc i\kern-.025em b}\kern-.08em
    T\kern-.1667em\lower.7ex\hbox{E}\kern-.125emX}}
\begin{document}

\title{Cardiorespiratory coupling improves cardiac pumping efficiency in heart failure}

\author{\IEEEauthorblockN{1\textsuperscript{st} Josh Border}
\IEEEauthorblockA{\textit{Department of Physics} \\
\textit{University of Bath}\\
Bath, United Kingdom}
\and
\IEEEauthorblockN{2\textsuperscript{nd} Andrew Lefevre}
\IEEEauthorblockA{\textit{Department of Physics} \\
\textit{University of Bath}\\
Bath, United Kingdom}
\and
\IEEEauthorblockN{3\textsuperscript{rd} Vishal Jain}
\IEEEauthorblockA{\textit{Department of Physics} \\
\textit{University of Bath}\\
Bath, United Kingdom}
\and
\IEEEauthorblockN{4\textsuperscript{th} Alain Nogaret}
\IEEEauthorblockA{\textit{Department of Physics} \\
\textit{University of Bath}\\
Bath, United Kingdom \\
email: A.R.Nogaret@bath.ac.uk \\
}
}

\maketitle

\begin{abstract}
Recent trials of a neuronal pacemaker have shown that cardiac pumping efficiency increases when respiratory sinus arrhythmia (RSA) is artificially restored in animal models of heart failure.  This novel device sheds new light on the functional role of RSA, which has long been debated, by allowing the strength of cardiorespiratory coupling to be artificially varied.  Here we show that RSA minimizes the cardiac power dissipated within the cardiovascular network.  The cardiorespiratory system is found to exhibit mode-locked synchronized regions within which viscoelastic dissipation is reduced relative to the scenario where cardiorespiratory coupling is absent.  We determine the gain in cardiac output as the magnitude of RSA increases.  We find that cardiac pumping efficiency improves up and until the cardiac frequency, within each breadth intake, is approximately 1.5 times greater than the cardiac frequency in the expiratory phase, at which point it reaches a plateau.  RSA was found to be most effective at low cardiac frequencies, in good agreement with clinical evidence.  Simulation of the cardiac power saved under RSA is in good agreement with the 17-20\% increase in cardiac output observed in RSA-paced animal models.
\end{abstract}

\begin{IEEEkeywords}
Neuronal pacemaker, respiratory sinus arrhythmia, cardiac output, synchronization, central pattern generator \footnote{Copyright: $979-8-3503-9205-0/24/\$31.00$ \textcopyright 2024 IEEE }
\end{IEEEkeywords}

\section{Introduction}
RSA is a physiological process by which respiration modulates heart rate, causing the heart to beat a little faster during a breadth intake than during the expiratory phase.  RSA has long been known to emanate from the modulation of vagal tone to the heart by brainstem central pattern generators~\cite{Anrep1936}.  However and in spite of its pervasiveness across Evolution, the functional significance of RSA has remained controversial.  One hypothesis is that it facilitates pulmonary gas exchange~\cite{Nicholas2003,BenTal2012}.  Analysis of cardiac oscillations provides evidence that it introduces a degree of synchronization between cardiac and respiratory rhythms~\cite{Skytioti2022,Schafer1999,Rzeczinski2002}.  The loss of RSA is a prognosis for the onset of heart failure~\cite{Hayano1996}.  In recent years, neuronal pacemakers have been developed that restore RSA by replicating the nonlinear response of brainstem circuits with silicon central pattern generators~\cite{Nogaret2021,Zhao2015,Nogaret2013}.  Remarkably, rodents~\cite{Callaghan2019} and large animal models of heart failure~\cite{Shanks2022,Riesenhuber2023} paced with a neuronal pacemaker have consistently demonstrated an increase in cardiac output in the 17-20\% range which we seek to interpret here on the basis of first principle simulations.  The ability to control the RSA dose administered allows us to accurately estimate the energy expended by the cardiorespiratory system in the presence of RSA relative to no RSA.  Our key results are that RSA reduces viscoelastic power dissipation in the vascular system by an amount comparable to the gain in cardiac output observed in animal trials.  This energy gain occurs in regions of mode-locked synchronization and by an amount which we determine.

\section{Method}

\begin{figure}[htbp]
\centerline{\includegraphics[width=\linewidth]{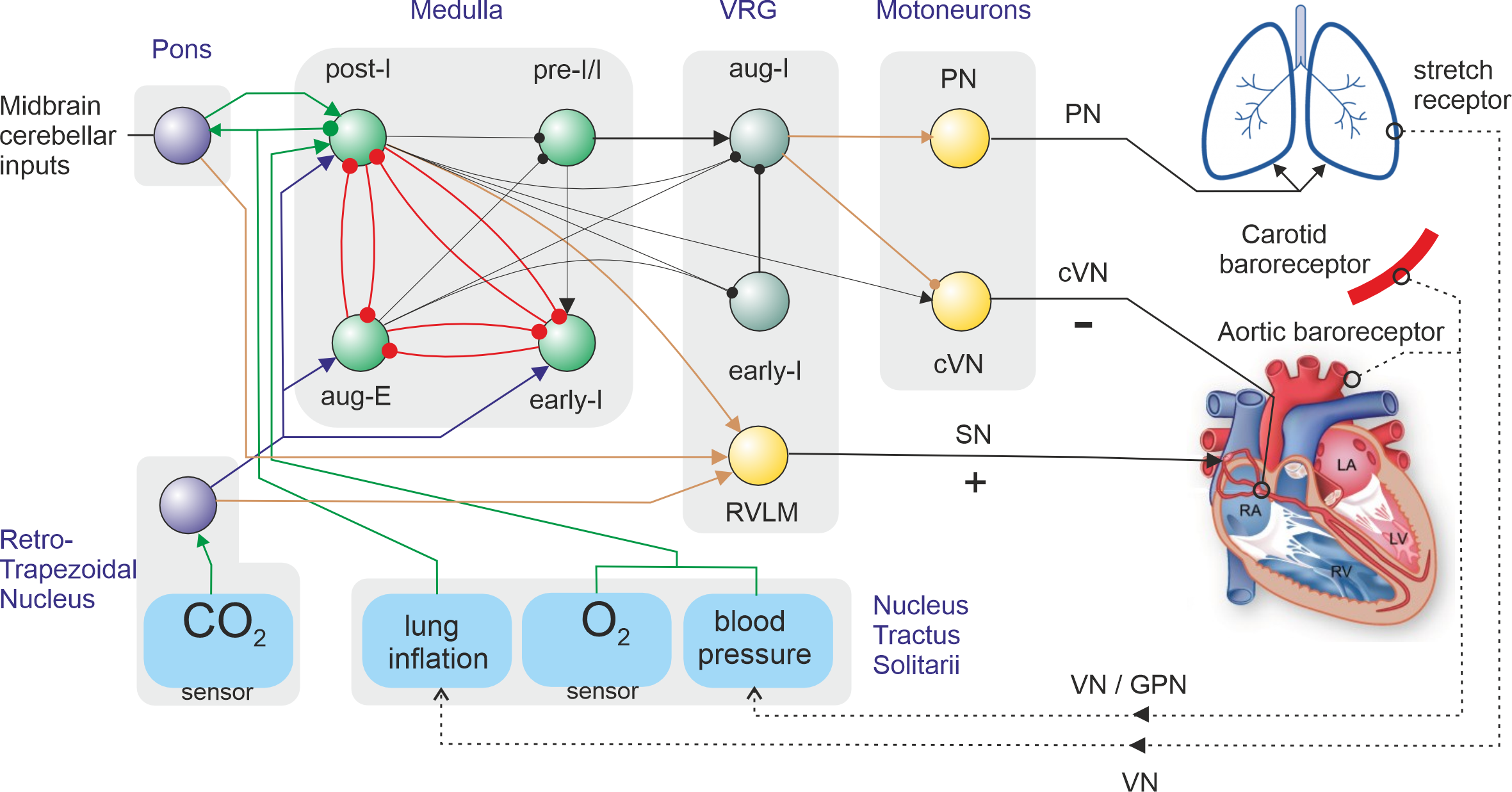}}
\caption{The respiratory central pattern generator (early-I, post-I, aug-E neurons) modulates vagal tone to the heart (cVN) based on physiological feedback from lung stretch receptors, arterial pO$_2$ and pCO$_2$, and baroreceptors.}
\label{fig:fig1}
\end{figure}

\begin{figure}[htbp]
\centerline{\includegraphics[width=\linewidth]{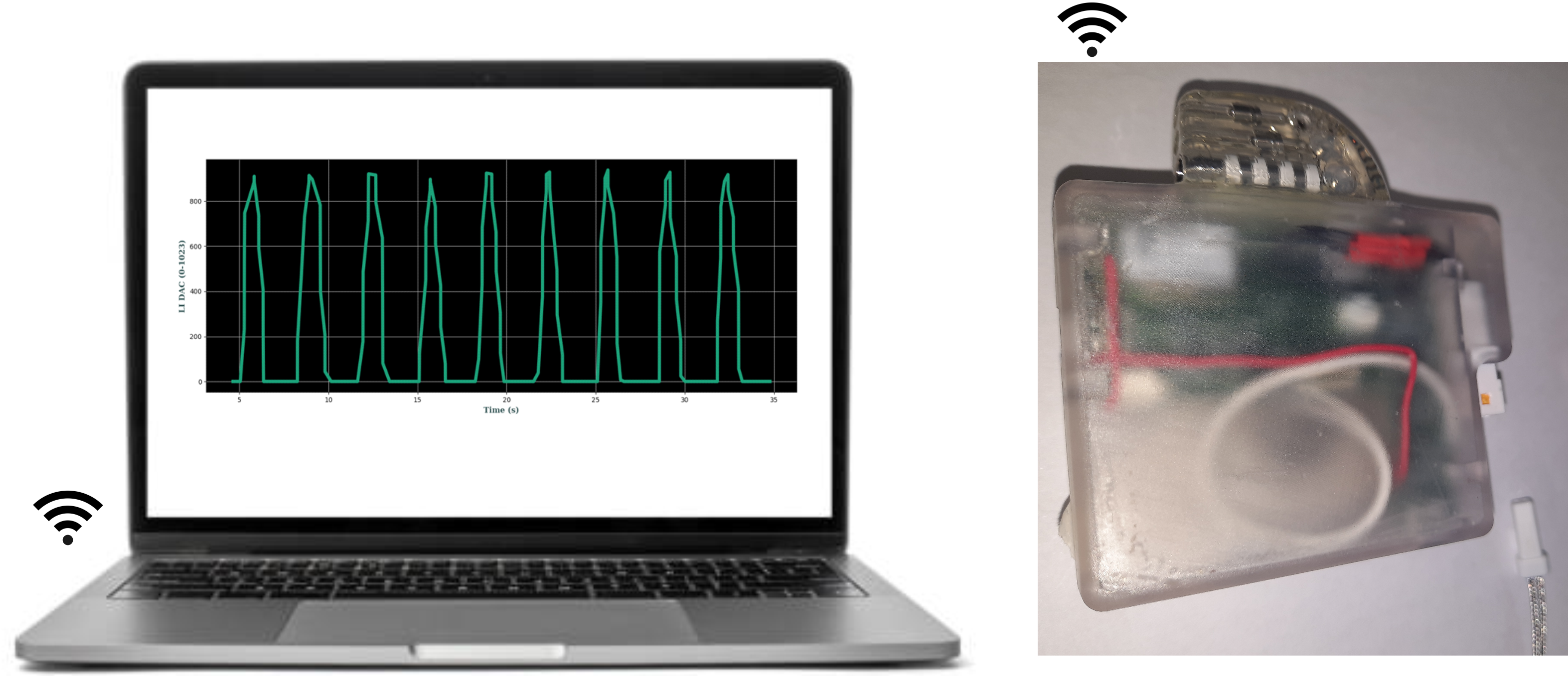}}
\caption{Our neuronal pacemaker delivers electrical pulses to the heart which are modulated on a beat-to-beat basis by the lung inflation signal (white connector).  The computer monitor displays the lung inflation signal received by the pacemaker.  Each pulse represents one breadth intake.}
\label{fig:fig2}
\end{figure}

The sequential discharge of three neuron populations (early-I, post-I, aug-E) in Fig.\ref{fig:fig1} in the brainstem produces the base respiratory rhythm.  In the process, the firing of the post-inspiratory (post-I) neuron increases vagal outflow to the heart (cVN).  This lowers heart rate at the onset of the expiratory phase and produces the RSA modulation seen in healthy individuals.  The neuronal pacemaker we have built (Fig.\ref{fig:fig2}) implements this circuit in silicon neurons to deliver electrical pulses to the heart that are modulated on a beat-to-beat basis by lung inflation.  The device was trialled on sheep and rats~\cite{Callaghan2019,Shanks2022,Riesenhuber2023} where it was found to increase cardiac output by ~19\% at the typical RSA dose of 13\% - i.e. when the ratio of cardiac frequency in the inspiratory phase relative to the expiratory phase was 1.13.
We computed the cardiac power saved in RSA relative to no RSA by first simulating the timing of cardiac pulses output by the pacemaker for a given respiratory pattern then by computing the viscoelastic power dissipated in lung alveoli arterioles that stretch and relax under both inhalation and systolic blood pressure.

\section{Results}

\begin{figure}[htbp]
\centerline{\includegraphics[width=\linewidth]{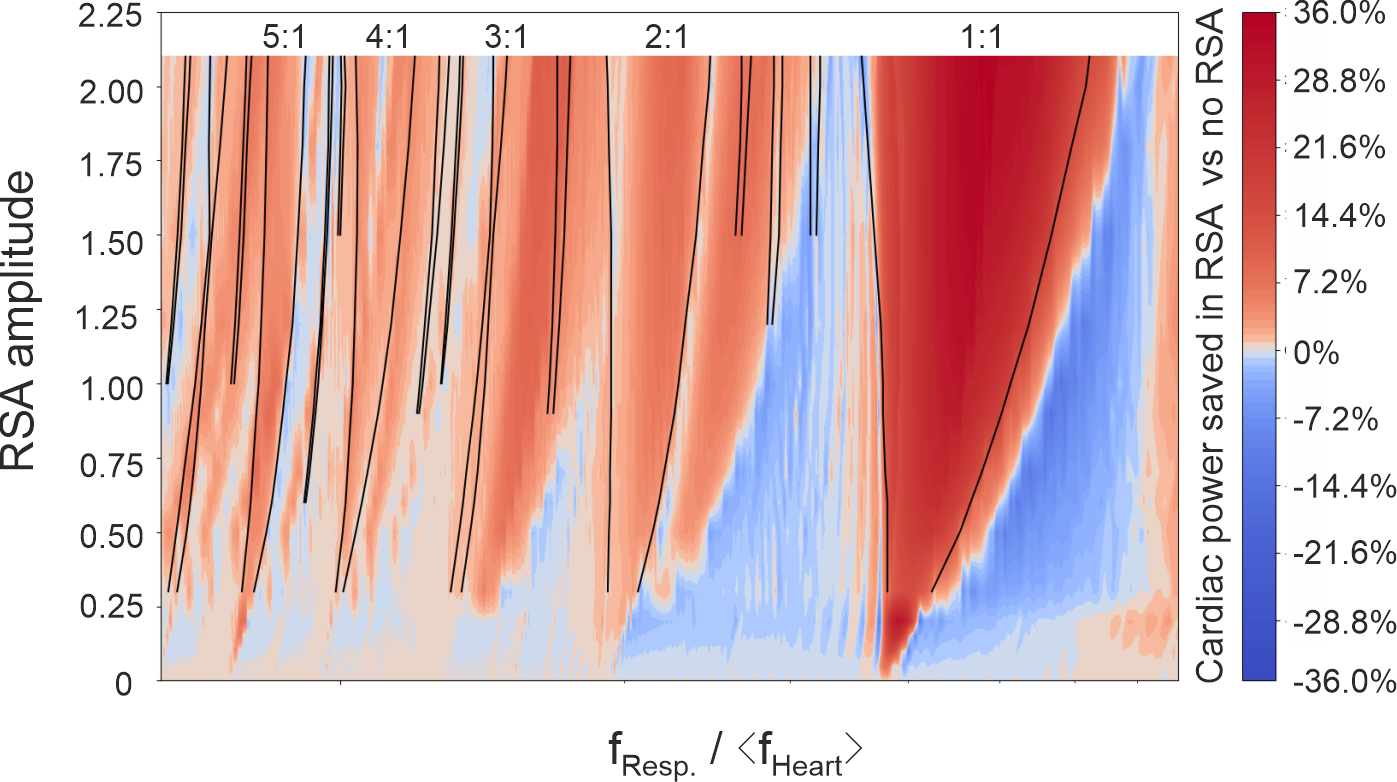}}
\caption{Colour map of the cardiac power saved in the presence of RSA relative to no RSA.  Mode-locked synchronization of cardiac and respiratory rhythms occurs as the mean cardiac frequency, $\langle f_{Heart} \rangle$, changes relative to the breathing frequency, $f_{Resp.}$.  These are the Arnold tongues $(m:n)=(1:1),(3:2),(2:1)\dots$ delimited by the black lines where $m$ $R-R$ intervals exactly match $n$ respiration periods.   The red (resp. blue) regions indicate a reduction (resp. increase) in viscoelastic power dissipated by the cardiorespiratory system with RSA relative to no RSA.  These regions become wider as the RSA coupling strength increases.  Gains in cardiac power efficiency saturate once the amplitude of RSA reaches 50\%.}
\label{fig:fig3}
\end{figure}

The nonlinear response of neurons in the central pattern generator produces a bias to synchronization between the respiratory and cardiac rhythms.  Whenever the cardiac and respiratory intervals are commensurate with each other, mode-locked synchronization ensues.  This is seen in Fig.\ref{fig:fig3} where Arnold tongues ($m:n$) form when $m$ cardiac oscillations exactly span $n$ breathing cycles.  Further additional narrow bands of synchronization arise from commensurability of fast inspiratory (resp. slow expiratory) cardiac frequencies to the inspiratory (resp. expiratory) intervals.  Having computed the timings of cardiac contractions in RSA, we proceeded to calculate the viscoelastic power dissipated in the presence of RSA normalised by the viscoelastic power dissipated without RSA.  We took great care to set the metronomic frequency of the unmodulated heart to the mean frequency the RSA-modulated heart $\langle f_{Heart}\rangle$ to avoid differences in power associated with trivial differences in cardiac frequencies.  Viscoelastic energy loss in the cardiorespiratory system occurs in arterioles surrounding pulmonary alveoli.  The periodic stretching and contraction of these vessels is the major source of impedance loading to the heart.  We calculated the viscoelastic power dissipated through this process, in RSA vs no RSA, in Fig.\ref{fig:fig3}.  There is clear evidence that within Arnold tongue regions, RSA minimizes viscoelastic dissipation by up to 25\% in the $1:1$ band and ~19\% in the $3:1$ band.  The latter best relates to cardiac/respiration frequency ratios in mammals.

\section{Discussion}

These results are in excellent agrement with the clinical observation that RSA is most effective in patients at rest (low heart rate).  In addition, energy gains of 19\% in the $3:1$ mode are in good agreement with the 17-20\% increase in cardiac output observed in rat and sheep models with reinstated RSA~\cite{Callaghan2019,Shanks2022}.  By considering the nonlinear response of brainstem neurons and the viscoelastic interactions between vascular and respiratory systems, we have successfully predicted gains in cardiac output induced by heart rate variability.

\section*{Acknowledgment}

This work was supported by the European Commission H2020 programme under project 732170.


\end{document}